\documentclass[twocolappendix]{emulateapj}

\usepackage[usenames,dvipsnames]{color}
\usepackage{comment}
\usepackage{graphicx}

\shorttitle{Disk Collisions} 
\shortauthors{McDowell, Duffell \& Kasen}

\begin{document}

\title{Interaction of a Supernova with a Circumstellar Disk}

\author{Austin T. McDowell, Paul C. Duffell and Daniel Kasen}
\affil{Astronomy Department and Theoretical Astrophysics Center, University of California, Berkeley, CA 94720}
\email{austintmcdowell@berkeley.edu}

\begin{abstract}

Interaction between supernova (SN) ejecta and a dense circumstellar medium (CSM) can power a luminous light curve and create narrow emission lines in the spectra. While  theoretical 
studies of interaction often assume a spherically symmetric CSM, there are observational indications that the gas surrounding 
some SN has a disk-like geometry.
Here, we use moving-mesh hydrodynamics simulations to study  the  interaction of  a SN with a disk and determine how the dynamics
and observable signatures may depend on the disk mass, thickness, and radial extent. We find that simple 
modifications to standard spherically-symmetric scaling laws
can be used to describe the propagation and heating rate of the interaction shock.
 We use the resulting shock heating rates to derive approximate  bolometric light curves, and provide
 analytic formulas that can be used to generate simple synthetic light curves for general supernova-disk interactions.
 For certain disk parameters and explosion energies, we are able to produce luminosities akin to those seen in super-luminous SN.
 Because the SN ejecta can flow around and engulf the CSM disk, the interaction region may become embedded and from certain viewing angles the 
 narrow emission lines indicative of interaction may be hidden. 
\end{abstract}


\section{Introduction} \label{sec:intro}

Observations of stellar outbursts such as those from $\eta$-Carinae \citep{Davidson_1997} and events like SN~2009ip \citep{Smith_2010, Foley_2011, Mauerhan_2013, Pastorello_2013, Margutti_2014} demonstrate that some massive stars undergo occasional episodes of extreme  mass loss.  If a star explodes as a supernova (SN)  shortly after such an eruption, the collision of the SN ejecta with previously ejected mass can thermalize the ejecta kinetic energy and power a luminous light curve.  The spectra of Type~IIn \citep{Filippenko_1997} and Type~Ibn SN \citep{Pastorello_2008} show narrow emission lines, indicative of interaction between the SN ejecta and a slow-moving, dense circumstellar medium (CSM). 




Previous numerical studies of interacting SN have typically assumed a spherical CSM distribution \citep[e.g.][]{1982ApJ...258..790C, 2010MNRAS.407.2305V, 2016ApJ...823..100H}. The ejected mass distribution in  $\eta$-Carinae, however, is clearly aspherical. In addition, spectropolarimetry of SN~2012ip and other interacting SN has been interpreted as evidence of a disk-like CSM geometry \citep[e.g.,][]{2014MNRAS.442.1166M} . An axisymmetric CSM could be the product of binary interaction; for example, the motion of a companion star embedded in the envelope of an inflated star  could drive mass ejection primarily in the equatorial plane \citep{Chevalier_2012,Pejcha_2016}.


Asymmetry of the CSM would have several significant consequence on the observable
properties of interacting SN. The aspherical interaction will thermalize only a fraction of the ejecta kinetic energy, affecting the luminosity and leading to viewing angle dependence of the light curves. Such effects have been explored in the 2D radiation-hydrodynamics simulations of \cite{Vlasis_2016}. In addition, if the SN ejecta flows around the disk and engulfs it, the emission line signatures may be hidden from view, at least from certain viewing angles. It is possible that such "embedded interaction" could explain observations of luminous supernovae that lack narrow emission features \citep{Quimby_2011}

In this paper, we perform 2D calculations using the moving-mesh hydrodynamics code, JET, to model the interaction between SN ejecta and a CSM that is assumed to be distributed in a disk surrounding the progenitor.  We study the propagation of the interaction shock, and calculate the amount of shock heating for various disk masses and disk opening angles. 
We then use the heating rate to estimate the properties of the resulting supernova light curves.   We show that most of the numerical results can be well described by simple analytic scalings.


In \S\ref{sec:numerics} we detail the numerical methods used to model the interaction between the supernova ejecta and CSM. This includes an explanation of the relevant fluid equations and initial conditions. The dynamics of the supernova interaction with the CSM are estimated analytically and calculated numerically in \S\ref{sec:shock}.  Estimates of important physical quantities such as the shock heating rate are presented and explained in \S\ref{sec:results}. Additionally, we explain the dependence of these quantities on the aspect ratio and mass of the surrounding disk.  These results are used to estimate bolometric light curves in \S\ref{sec:obs}. Finally, in \S\ref{sec:disc} we summarize our results.


\section{Numerical Set-Up} \label{sec:numerics}

\begin{figure}
\epsscale{1.15}
\plotone{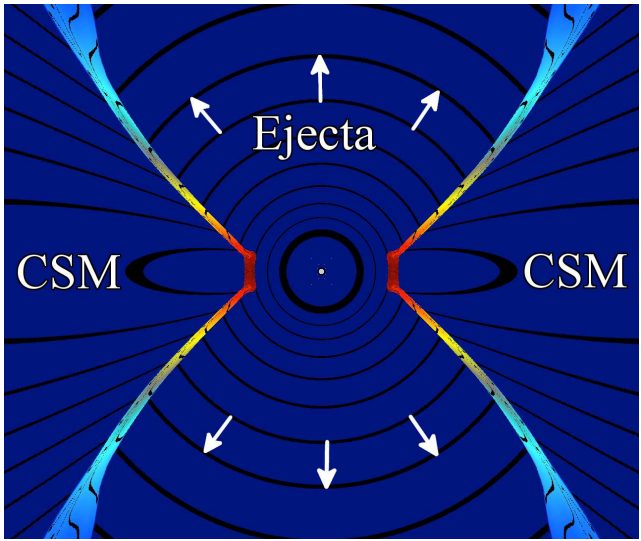}
\caption{ Possible distribution of CSM around SN explosion. Snapshot taken at time $t \approx 1$ day since explosion. Red areas correspond to regions of high pressure and heating. Contour lines depict log($\rho$). Ejecta expands freely along the north and south poles but collides with the CSM along the equator. Red regions along the equator are the sites of the strongest interaction and the greatest amount of heating. Ejecta to disk mass ratio is 100 with $M_{ej}=10M_{\odot}$ and $E_{ej}=10^{51}$ ergs. 
\label{fig:pretty} }
\end{figure}

\begin{figure*}
\epsscale{1.15}
\plotone{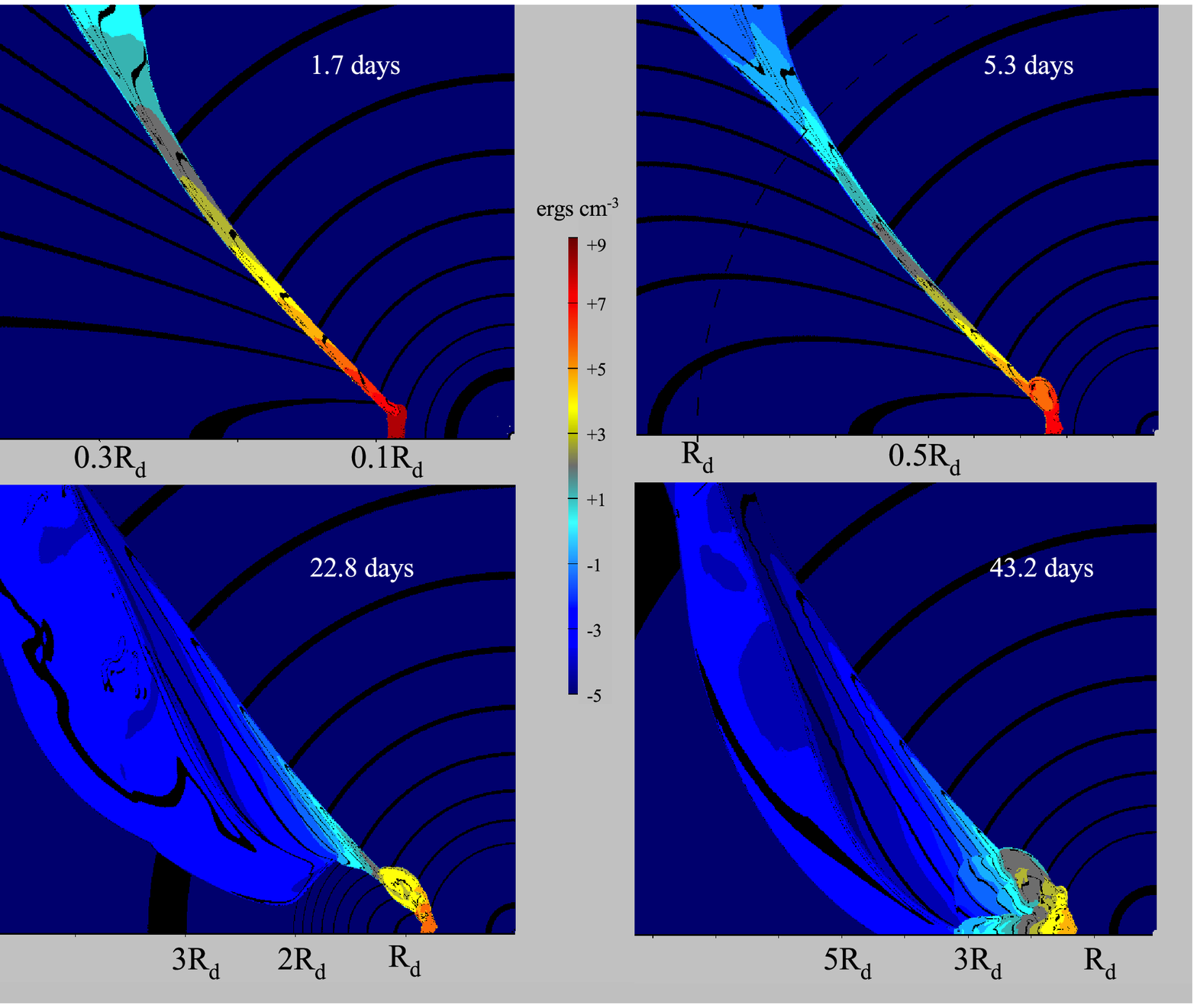}
\caption{ Evolution of CSM interaction. Snap shots are taken from the upper left quadrant of figure (\ref{fig:pretty}). Red and blue regions correspond to high and low pressure, respectively. Black contour lines show log(density). In the first two images we see high pressure, and therefor heating, concentrated in small regions around the equator. The third image shows the CSM beginning to wrap round the outer edge of the disk, and the last image shows when the CSM has engulfed the disk. For these snapshots we used a mass ratio of $M_{ej}/M_{d}=100$ with $M_{ej}=10M_{\odot}$, $E_{ej}=10^{51}$ ergs, and $R_d = 10^{15}$ cm.  
\label{fig:panel} }
\end{figure*}

We carried out numerical  calculations using  JET  \citep{2011ApJS..197...15D, 2013ApJ...775...87D}, a moving mesh hydrodynamics code that is effectively Lagrangian due to the radial motion of computational zones. The code integrates the equations of two-dimensional (2D) axisymmetric hydrodynamics

\begin{equation} 
\partial_t ( \rho ) + \nabla \cdot ( \rho \vec v ) = 0,
\end{equation} 
\begin{equation}
\partial_t ( \rho v_r ) + \nabla \cdot ( \rho v_r \vec v + P \hat r ) = ( 2 P + \rho v_\theta^2 )/ r,
\end{equation} 
\begin{equation}
\partial_t ( r \rho v_\theta ) + \nabla \cdot ( r \rho v_\theta \vec v + P \hat \theta ) = P \cot \theta,
\end{equation} 
\begin{equation}
\partial_t ( \frac12 \rho v^2 + \epsilon ) + \nabla \cdot ( ( \frac12 \rho v^2 + \epsilon + P ) \vec v ) = 0,
\end{equation} 
where $\rho$ is density, $P$ is pressure, $\epsilon$ is the internal energy density, and $\vec v$ is the velocity.  The equation of state is assumed to be that of a radiation dominated gas, $\epsilon = 3 P$.

The initial density configuration consists of outflowing  SN ejecta, a disk, and a surrounding low-density wind

\begin{equation}
\rho(r,\theta, t) = \rho_{\rm ej}(r,\theta,t) + \rho_{d}(r,\theta) + \rho_{\rm wind}(r).
\end{equation} 
We describe the ejecta density with the broken power-law profile \citep{Chevalier_Soker_1989} commonly adopted for
core collapse SN

\begin{equation}
\label{eq:rhoej}
\rho_{\rm ej}(r,\theta,t) = \left\{ \begin{array}
				{l@{\quad \quad}l}
                                \rho_T \left(\frac{r}{v_0 t}\right)^{-d} & r < v_0 t \\
				\rho_T \left(\frac{r}{v_0 t}\right)^{-n} & v_0 t < r < v_{\rm max}  t	\\  
    			0 & r > v_{\rm max} t	\\ 
    			\end{array} \right.    
\end{equation}
where

\begin{equation}
\rho_T = \frac{(n-3)(3-d)}{4\pi(n-d)} \frac{M_{ej}}{(v_0 t)^3}
\end{equation}

\begin{equation}
v_0 = \left( 2\frac{E_{ej}}{M_{ej}} \frac{(5-d)(n-5)}{(3-d)(n-3)} \right)^{1/2}
\end{equation}
 where $M_{ej}$ is the ejecta mass and $E_{ej}$ is the ejecta energy. We take the
exponents to be $n=10$ and $d=1$ and the maximum ejecta velocity $v_{\rm max} = 10 v_0$. 

The disk density profile is a power law in radius which cuts off sharply at $r = R_{d}$

\begin{equation}
\label{eq:diskden}
\rho_{d}(r,\theta) = \alpha { M_{d} \over R_{d}^3 } \left( {r \over R_{d} } \right)^{-s} e^{2 (sin(\theta)-1) \over h^2 } e^{ -(r/R_{d})^4 }
\label{eq:rho_d}
\end{equation}
where 
\begin{equation}
\alpha = {\sqrt{ h^{-2} + \pi/4 } \over 10.2 },
\end{equation} $M_d$ is the disk mass, and $R_d$ is the outer disk radius. The dimensionless quantity $h$ describes the aspect ratio of the disk and is equal to the scale height divided by radius.   The exponential $\sin(\theta)-1$ dependence in equation (\ref{eq:rho_d}) arises from an assumption that the disk is isothermal and in pressure equilibrium. That is to say, the vertical pressure gradient is balanced by the vertical component of the gravitational force and integrating over this gradient returns a factor of $e^{\frac{2(sin(\theta)-1)}{h^2}}$ in the equation for density. The constant $\alpha$ provides the correct normalization in the limit of both thin and thick disks, while the factor $e^{-(r/R_d)^4}$ creates a smooth density cutoff at $r = R_d$. 

The wind density is that of a constant velocity outflow 
\begin{equation}
\rho_{\rm wind} = {A_{\rm wind} \over r^2}.
\end{equation}
where $A_{\rm wind}$ is a constant chosen to be sufficiently small ($\sim10^{-8}$ \text{g/cm}) so that the wind does not significantly affect the interaction between the ejecta and the disk.

Ultimately, we wish to measure the power deposited in the disk by shocks, so that we can estimate the luminosity that may be radiated in the SN light curve. The shock heating is not directly calculated by the code, but can be estimated indirectly by evolving the fluid entropy as a passive scalar (assuming no shocks), and comparing this to the true fluid entropy. Specifically, the shock heating over a time $dt$ can be written as
\begin{equation}
dQ = \int dV(d\epsilon - d\epsilon_{\rm isentropic})
\end{equation} 
where $\epsilon$ is the energy density and $\epsilon_{\rm isentropic}$ is what the energy density would be if entropy was conserved. Since $\epsilon = P/(\gamma-1)$, where $P$ is the pressure and $\gamma$ is the adiabatic index, we can write
\begin{equation}
\label{eq:qdotintegral}
dQ = \int dV\frac{(dP - dP_{\rm isentropic})}{\gamma -1}.
\end{equation}

Isentropic pressure can be expressed in terms of the density and entropy 
\begin{equation}
P_{\rm isentropic} = \rho^{\gamma}e^s.
\end{equation}
$P_{\rm isentropic}$ is calculated at each checkpoint by evolving the specific entropy $s$ as a passive scalar and resetting it to ${\rm ln}(P/\rho^{\gamma})$ at the end of each checkpoint. Then $\dot{Q}$ is calculated at the beginning of each checkpoint via equation (\ref{eq:qdotintegral}).


\section{Shock Dynamics} \label{sec:shock}

\subsection{Numerical Shock Dynamics}

Figure~\ref{fig:pretty} shows the structure of the ejecta/CSM interaction.
Heating takes place in a small region near the equator where the disk and ejecta collide and a strong shock forms. The disk diverts the ejecta flow, leading to an hourglass geometry.
Figure~\ref{fig:panel} shows a time series of the dynamics, where the ejecta is seen to eventually flow back around the disk edges, engulfing it. As the ejecta wraps around the disk, the viewing angle at which the interaction is visible begins to decrease. Eventually, the ejecta surrounds the whole disk and the interaction is hidden. This can be seen in the final panel of Figure \ref{fig:panel}.  
When the mass of interacting ejecta has become comparable to the disk mass, the entire disk is accelerated and swept up with the ejecta. At this point the shock weakens, the heating drops rapidly, and any narrow lines from the interaction will have been broadened. Additionally, in Figure 2 it is clear that the interaction is subject to fluid instabilities.  The Rayleigh-Taylor instability is present in the shocked region between ejecta and CSM, due to the jump in density between ejecta and CSM (this instability would also be present for a spherically symmetric initial condition).  Additionally, the shear at the surface of the disk appears to generate Kelvin-Helmholtz instability, which is manifested in Figure 2 as a large eddy in the second and third panels.


\subsection{Shock Propagation: Analytic Scalings} \label{sec:shockscalings}
Features of the interaction dynamics, in particular the position of the equatorial shock  as a function of time, $R_s(t)$, can be understood using analytic scaling arguments.  We first consider the regime in which the disk mass is much smaller than the ejecta mass, $M_{\rm ej} \gg M_{\rm d}$. We assume the ejecta density profile is described by the power-law in the outer region ($ v_0 t < r < v_{\rm max} t$) of equation (\ref{eq:rhoej}), and the disk density is given by the thin-disk limit of equation (\ref{eq:diskden}) with $\theta=\pi/2$. The evolution of $R_s(t)$ is then self-similar and 
its value  can be determined up to a constant by finding where the ejecta density equals the CSM density, giving
\begin{equation}
R_s(t) \sim \left({ M_{\rm ej} \over M_{d}} h (E_{\rm ej}/M_{\rm ej})^{(n-3)/2} { t^{n-3} \over R_{d}^{s-3} } \right)^{1/(n-s)}.
\end{equation}
\begin{equation}
\label{eq:chevrs}
\sim R_{d} \left({ M_{\rm ej} \over M_{d}} h \right)^{1/(n-s)} \left( { v_0 t \over R_{d} } \right)^{(n-3)/(n-s)},
\end{equation}

This scaling  $R_s(t) \sim t^{(n-3)/(n-s)}$ is the 1D self-similar solution of \cite{1982ApJ...258..790C}  for ejecta colliding with a spherical medium.  It should not be surprising that a self-similar solution exists in 2D, as no new scales are introduced until the shock reaches $R_{d}$. This scaling should be valid so long as the swept-up disk mass is sufficiently smaller than the ejecta mass. 
In our example, $n=10$ and $s=2$, which gives $R_s(t) \propto t^{7/8}$.

 In the opposite regime, where $M_{\rm ej} \ll M_{\rm d}$, one expects a Sedov-Taylor scaling.  This can be found by neglecting the ejecta mass and equating the velocity of the shock with the ratio of energy to swept-up mass
\begin{equation}
v_s^2 \sim \left[ \frac{R_s(t)}{t} \right]^2 \sim \frac{E_{\rm ej}}{ \rho(R_s) R_s^3 }.
\end{equation}
which leads to the following scaling for the shock radius
\begin{equation}
\label{eq:sedtay}
R_s(t) \sim \left( { E_{\rm ej} ~h ~t^2 \over M_{d} } R_{d}^{3-s} \right)^{1/(5-s)}.
\end{equation}
For our value of $s=2$ this gives
\begin{equation}
R_s(t) \propto t^{2/3}
\end{equation}
Below we show that these analytic scalings well describe the detailed numerical results.

\subsection{Shock propagation: Numerical Results}
\begin{figure}
\epsscale{1.15}
\plotone{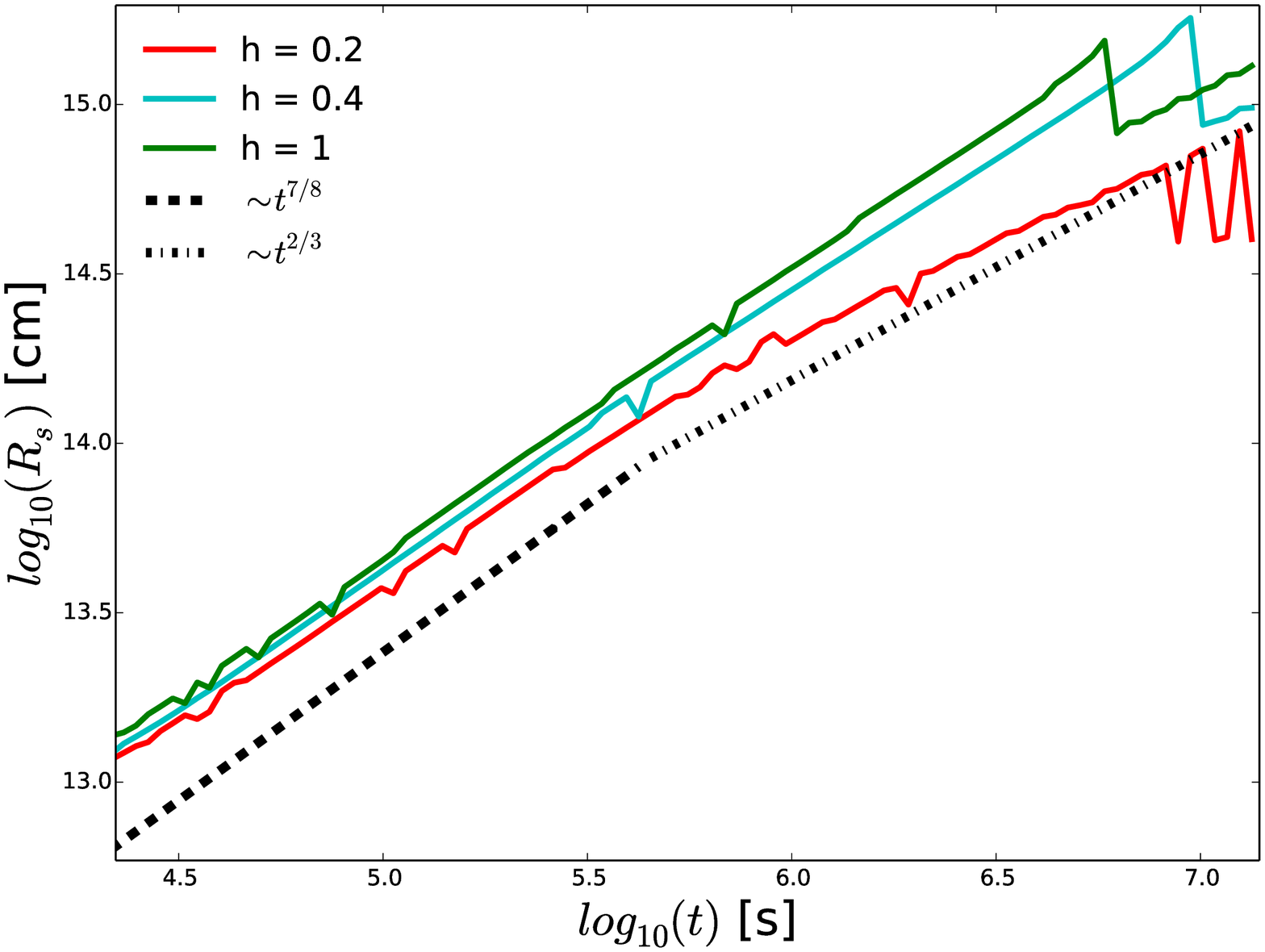}
\caption{ Position of the forward shock front as a function of time when the disk mass is equal to ejecta mass, $M_{ej}=10M_{\odot}$, and $E_{ej}=10^{51}$ ergs. Solid lines show the shock position found from the numerical models while dashed lines show analytic shock position found using equation (\ref{eq:chevrs}) and dot-dashed lines show analytic shock position found using equation (\ref{eq:sedtay}). Both analytic lines are offset by a factor of 1.5. The transition between the two scalings occurs as the swept up disk mass becomes comparable to the mass of the ejecta that interacts with the disk. 
\label{fig:rs_v_t_mrat1} }
\end{figure}

Figure \ref{fig:rs_v_t_mrat1} shows the forward shock position for the case of a disk mass equal to the ejecta mass. The analytic scalings of \S\ref{sec:shockscalings} well describe the shock propagation.
We expect the evolution should transition from being described by equation (\ref{eq:chevrs}) to more closely following equation (\ref{eq:sedtay}) at a time that depends on $h$. 

In the low disk mass limit, equation (\ref{eq:chevrs}) says that the shock position should scale as $R_s \propto h^{1/(n-s)}$, or $R_s \propto h^{1/8}$ for our fiducial values $n=10, s=2$. 
We find a similar scaling at early times in the numerical results shown in  Figure \ref{fig:rs_v_t_mrat1}. 
From the figure we can also see that the transition towards the Sedov-Taylor solution happens earlier for narrower disks, i.e. a smaller value of $h$. The transition time occurs when the swept up disk mass is comparable to the amount of ejecta mass that interacts with the disk. The amount of interacting mass is proportional to the disk's opening angle and thus $h$.
This means that as $h$ is decreased, less disk mass needs to be swept up before the transition between Chevalier and Sedov-Taylor scalings will occur. After this transition, equation (\ref{eq:sedtay}) predicts that the shock position will scale as $R_s \propto h^{1/3}$. Figure \ref{fig:rs_v_t_mrat1}, shows that the late-time behavior does not follow the Sedov-Taylor scaling as closely as the early-time behavior does the Chevalier solution, suggesting that a more massive disk is needed for the Sedov-Taylor scaling to reasonably approximate the solution at the times shown.  

Our hydrodynamical calculations neglect the effects of radiation transport, which may influence of the dynamics. In the limit that the optical depth in the interaction region is low enough that radiation can effectively escapes, we expect the dynamical behavior to approach that of a gamma $\approx$ 1 fluid. The analytic scaling relations for the shock position, which are independent of gamma,  may continue to hold to reasonable approximation, but the structure of the shocked gas will be different. Due to radiative cooling, the post shocked gas is likely to form a much cooler, thinner and denser shell, which may lack sufficient pressure for lateral forces to cause the gas to flow back around the disk and engulf it.  Fully coupled radiation hydrodynamical calculation will be needed to accurately capture the evolution, however since we are interested here primarily in scenarios in which the radiative diffusion time is $\gtrsim$ the dynamical time,  we expect our numerical hydrodynamical results to provide a fair approximation to the basic character of disk interaction




\section{Heating Rate from Interaction} \label{sec:results}

The heating rate resulting from  interaction shocks is an important quantity, as it determines the rate at which the ejecta kinetic energy is transformed into thermal
energy that can be radiated in the supernova light curve.  
We can analytically estimate the power produced by the shock as
\begin{equation}
\dot Q \sim P_{\rm shock} v_{\rm shock} A_{\rm shock},
\end{equation}
where $P_{\rm shock}$ is the pressure behind the shock front, $v_{\rm shock}$ is the shock velocity (which can be calculated from equation (\ref{eq:chevrs})) and $A_{\rm shock}$ is the surface area of the shock.
For highly supersonic shocks, $P_{\rm shock}$ is  given by 

\begin{equation}
P_{\rm shock} \sim \rho_{d}(R_s) v_{\rm shock}^2.
\end{equation}
The shock velocity scales as $v_s \sim R_s(t)/t$, 
and the shock surface area is $A_{\rm shock} \sim h R_s^2$.
Putting these together gives

\begin{equation}
\dot Q \sim \rho_{d}( R_s ) \frac{h R_s^5}{t^3}.
\end{equation}
The equatorial disk density profile for a thin disk (equation (\ref{eq:diskden}) with $\theta = \pi/2$ and $h \ll 1$) is
\begin{displaymath}
\rho_{d}(R_s) \sim h^{-1}\frac{M_d}{R_d^3} \left(\frac{R_s}{R_d}\right)^{-s} .
\end{displaymath}
Using this and $R_s(t)$ given by equation (\ref{eq:chevrs}), we get the heating rate

\begin{equation}
\dot Q \sim M_d R_d^2 \left(\frac{v_0}{R_d}\right)^{\frac{(5-s)(n-3)}{n-s}}\left(\frac{M_{ej}}{M_{d}}h\right)^{5-s \over n-s}t^{{(5-s) (n-3) \over n-s} - 3}
\end{equation}
For the model parameters used here, $n = 10$, $s=2$, this gives a time dependence of $\dot Q \sim t^{-3/8}$. 

The shock heating rate can thus be simplified and written as

\begin{equation}
\label{eq:anheating}
\dot Q = \left\{ \begin{array}
				{l@{\quad \quad}l}
				{\frac{M_d}{R_d} \left(\frac{E_{ej}}{M_{ej}}\right)^{3/2}} \left(\frac{M_{ej}}{M_{d}} h\right)^{3/8} (t/t_0)^{-3/8} & t < t_{\rm sweep} 	\\  
    			0 & t > t_{\rm sweep}	\\ 
    			\end{array} \right.    
\end{equation}

where

\begin{equation}
t_0 = R_d/v_0
\end{equation}
is the characteristic timescale for the ejecta to reach the outer disk radius.

\begin{figure}
\epsscale{1.15}
\plotone{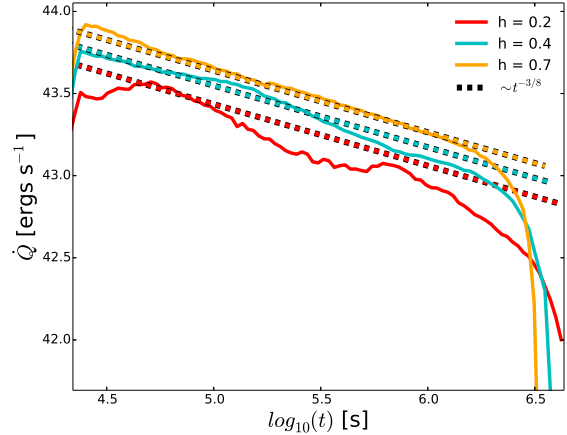}
\caption{ Heating rate as function of time for ejecta interacting with a CSM disk of mass of $M_{ej}/10$, where $M_{ej}=10M_{\odot}$ and $E_{ej}=10^{51}$ ergs. Solid lines show values calculated using equation (\ref{eq:qdotintegral}) and dotted lines show analytic curves given by equation (\ref{eq:anheating}).
\label{fig:qdot_mrat10} }
\end{figure}

In Figure (\ref{fig:qdot_mrat10}) we compare this analytic scaling to the heating rate from our numerical calculations. We find that the time dependence is reasonably close to the $t^{-3/8}$ law predicted, while the dependence on $h$ may be stronger than predicted. There may be an additional dependence on $h$ which encodes how difficult it is for the shock energy to flow around the disk.  If the energy is allowed to escape around the disk, this reduces the shock strength and heating rate. 

The self-similar ejecta and disk dynamics assumed in the above analysis will persist only until a time  $t_{\rm sweep}$ when the shock has passed through the entire disk.  Assuming $M_{d} \ll M_{\rm ej}$, this time is given by setting $R_s$ equal to $R_{d}$ in equation (\ref{eq:chevrs})
\begin{equation}
\label{eq:tsweep}
t_{\rm sweep} \sim {R_{d} \over v_0} \left( {{ M_{\rm ej} \over M_{d}} h} \right)^{ -1 \over n-3 }.
\end{equation}
For a steep ejecta gradient, $n \gg 1$, this essentially reduces to the crossing time $R_{d} \over v_0$ with very weak dependence on the disk properties. 

In the opposite regime where $M_{d} \gg M_{\rm ej}$ one can set equation (\ref{eq:sedtay}) equal to $R_{d}$ and recover
\begin{equation}
t_{\rm sweep} \sim \frac{R_{d}}{v_0} \left( \frac{M_{ej}}{M_{d}} h \right)^{-1/2}
\end{equation}
In this case, the more massive disk slows the shock and the sweep-up time depends more strongly on the disk properties.


\begin{figure}
\epsscale{1.15}
\plotone{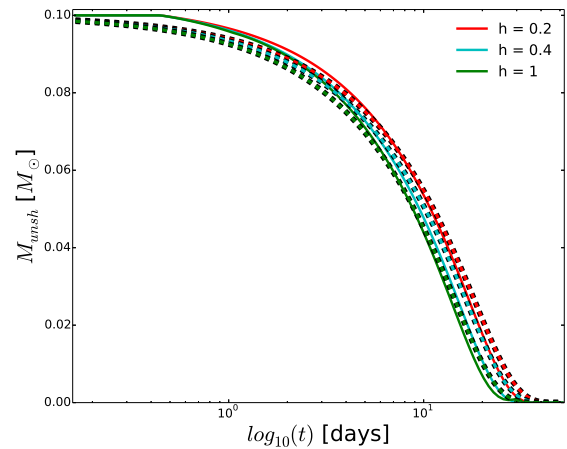}
\caption{ Unshocked disk mass as a function of time for $M_{ej}/M_{d}=100$, $M_{ej}=10M_{\odot}$, and $E_{ej}=10^{51}$ ergs. Solid lines show numerical values while dashed lines show analytic curves found by integrating the total mass between the shock radius and infinity. 
\label{fig:unshockedmass_mrat100} }
\end{figure}


\section{Observational Signatures} \label{sec:obs}

The energy dissipated by the interaction shock will be radiated, potentially leading to a luminous supernova light curve. Detailed predictions of the observable signatures require multi-dimensional radiative transfer calculations; here we simply provide  estimates of some of the basic behaviors.

Given that the CSM typically moves slowly compared to the SN ejecta, a signature of interaction
is the presence of narrow emission lines in the spectrum. However, as the interaction shock travels through the disk, the CSM is accelerated to SN-like velocities. Once the mass of unshocked disk material becomes small, any CSM emission lines should be significantly broadened and will no longer be indicators of interaction.

Figure \ref{fig:unshockedmass_mrat100} shows the unshocked disk mass over time for numerical calculations with $M_{\rm ej}/M_d = 100$ and various values of $h$.
 The unshocked disk mass can also be estimated analytically by integrating the total disk mass between $R_s(t)$ and infinity, and this analytic result is also plotted in Figure \ref{fig:unshockedmass_mrat100}.  The unshocked disk mass is seen to initially decline gradually with time, then fall of rapidly at $t \approx t_{\rm sweep}$. This sweep-up time, which
 depends  only weakly on $h$ (equation (\ref{eq:tsweep})) provides an estimate for the latest time one might expect narrow emission lines to be visible.

The bolometric luminosity of the interacting SN will depend on the heating rate $\dot Q$.  Since the ejecta and CSM are likely to be optically thick at early times, an estimate of the
light curve must take into account radiative diffusion.    As a rough model, we assume that there is a time, $t_{sn}$, where the opacity of the material becomes low enough for the radiation to escape. This timescale is typically approximated by \citep{Colgate_McKee_69, Arnett_1979}
\begin{equation}
t_{sn} \sim \kappa^{1/2} M^{3/4} E_{\rm ej}^{-1/4}
\end{equation}
Where the opacity $\kappa$ is typically of order the value for electron scattering, and the effective diffusion mass $M$ may include both the disk and the ejecta mass.

Given a diffusion timescale, $t_{\rm sn}$, and the heating rate $\dot Q(t)$, one can generate a bolometric light curve using the  integral \citep{Arnett_1982}
\begin{equation}
\label{eq:lcintegral}
L_{sn}(t) = e^{-(t/2t_{sn})^2}\int_0^t \dot{Q} e^{(t'/2t_{sn})^2}(t'/t_{sn})dt'
\end{equation}
where $\dot{Q}$ is either given by equation (\ref{eq:anheating}) or calculated numerically using fluid properties as discussed in section \ref{sec:results}.  
Similar light curve calculations have been discussed in \cite{Chatz_2012} and \cite{Moriya_2013}.
This method is only intended to approximate the effects of radiative diffusion; more detailed light curve prediction  require full radiation transfer calculations \citep{Vlasis_2016}.


Figure (\ref{fig:lc_mrat100_02}) shows both analytic and numeric light curves for $h=0.2$, $M_{ej}/M_{d}=100$, and 
 different values of $t_{sn}$ in the range 1-4 days.  The model light curves show the expected dependence on $t_{sn}$; smaller values of $t_{sn}$ allow for more of the radiation to escape and increase the overall luminosity. The light curves using the analytic heating rate agree reasonably well with those based on the numerical heating rates. 

Figure (\ref{fig:lc_mrat1}) shows calculated light curves for $h=0.2$ and a larger relative disk mass, $M_{ej}/M_{d}=1$. The larger disk mass means that a greater fraction of the SN ejecta is decelerated and the light curves are brighter.  In this regime, the assumptions made in deriving an expression for $\dot{Q}$ no longer hold and the agreement between the analytic and numeric curves starts to falter. However, when $M_{ej}=M_{d}$ we are able to reach peak luminosities similar to those seen in super-luminous SN.


\begin{figure}
\epsscale{1.15}
\label{fig:lc_mrat100_02}
\plotone{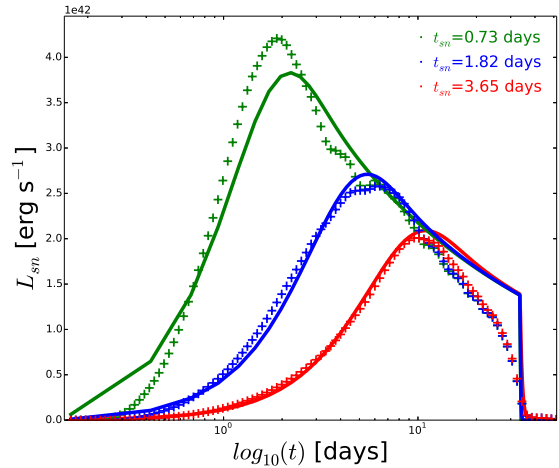}
\caption{ Lightcurves calculated for $h=0.2$, $M_{ej}/M_{d}=100$, and $M_{ej}=10M_{\odot}$. Colored crosses show lightcurves calculated using numerical heating rate and lines show lightcurves found using an analytic heating rate. Different colors correspond to different values of $t_{sn}$. For each curve, an ejecta energy of $10^{51}$ ergs is used.   
\label{fig:lc_mrat100_02} }
\end{figure}

\begin{figure}
\epsscale{1.}
\label{fig:lc_mrat1}
\plotone{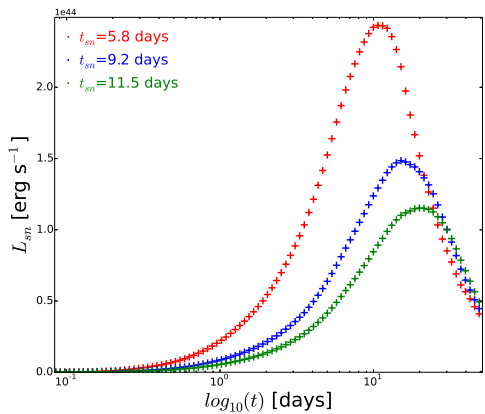}
\caption{ Lightcurves for $h=0.2$, $M_{ej}/M_{d}=1$, and $M_{ej}=10M_{\odot}$. Different colors correspond to different values of $t_{sn}$. For each curve we assume an ejecta energy of $10^{52}$ ergs.
\label{fig:lc_mrat1} }
\end{figure}


\section{Discussion and Conclusion} \label{sec:disc}

We have run numerical calculations of the interaction between a supernova and a non-spherical CSM.  We have found that the initial propagation of the  interaction shock can be reasonably estimated by the \cite{1982ApJ...258..790C} scalings which depend weakly on the CSM properties.  When the swept up CSM mass becomes comparable to the ejecta mass, the Sedov-Taylor scalings can be used, although there is not quite as close a match between analytic predictions and the multi-dimensional numerical results.
These scalings also give an accurate estimate for $t_{\rm sweep}$, the time when the shock has swept up the entire disk. 
For times  $t > t_{\rm sweep}$, the heating from interaction drops sharply and the any narrow emission lines will become Doppler-broadened to
SN-like velocities.


We calculated the rate of heating  produced as the interaction shock thermalizes the SN kinetic energy and used this to produce approximate bolometric light curves.  We also presented analytic estimates for the heating rate that roughly match the results of the numerical simulations.  Disagreement between the analytical and numerical heating rates is likely related to the complex multi-dimensional effects as the SN ejecta flows around the disk.  Our models reinforce the intuition that thicker and more massive disks will result in greater overall heating.  Although the solid angle subtended by the disk is proportional to $h$, the heating rate (for fixed disk mass) scales sub-linearly, approximately $\dot{Q} \propto h^{3/8}$. This is because increasing $h$ lowers the disk density, which weakens the shock heating. For example, confining the CSM to a narrow disk with $h = 0.1$ (opening angle $\approx 10^\circ$) produces a heating rate that is only $\approx 0.4$ times less than a corresponding spherical CSM of the same mass. 

Our approximate bolometric light curves  achieve higher luminosities for disks with more mass and larger opening angles. 
For low mass disks with relatively short diffusion times, we find quickly rising and fading light curves of moderate brightness. 
These may be of relevance to the recently discovered class of rapidly-evolving transients 
\citep{poznanski10, drout14, arcavi16}. For large disk masses ($M_{\rm d} \approx M_{\rm ej}$) and high explosion energies, the luminosity from interaction can approach that seen in the class of super-luminous supernovae \citep{GalYam12}.

An interesting consequence of  a disk-like CSM geometry is that the narrow emission lines typically used to diagnose interaction may, from certain viewing angles, be hidden from view. In the hydrodynamics simulations, the SN ejecta is diverted by the densest part of the disk and flows around it, embedding the primary region of shock heating. Once the fastest moving ejecta passes the outer radius of the disk, it can flow back around to engulf the disk entirely. We speculate that from certain viewing angles (those nearer the pole) the narrow emission lines produced in the slow-moving unshocked CSM  will be obscured by the optically thick SN ejecta. This leads to the possibility that some super-luminous SN may be powered by interaction even if they lack the characteristic narrow line signatures.  Multi-dimensional radiation transport calculations will be needed to accurately predict the spectral signatures.

We reiterate that the basic features of SN-disk interaction dynamics and light curves can be  estimated analytically. The important expressions are the shock position as a function of time
\begin{displaymath}
R_s(t) \sim \left( h {  M_{\rm ej} \over M_{d}} 
\left( \frac{E_{\rm ej}}{M_{\rm ej}} \right)^{(n-3)/2} { t^{n-3} \over R_{d}^{s-3} } \right)^{1/(n-s)}
\end{displaymath}
and the heating rate, which for typical ejecta density power-law exponents ($n=10, s=2$) is
\begin{displaymath}
\dot Q = \left\{ \begin{array}
				{l@{\quad \quad}l}
				{\frac{M_d}{R_d} \left(\frac{E_{ej}}{M_{ej}}\right)^{3/2}} \left(\frac{M_{ej}}{M_{d}} h\right)^{3/8} 
				\left( t/t_0 \right)^{-3/8} & t < t_{\rm sweep} 	\\    
    			0 & t > t_{\rm sweep}	\\ 
    			\end{array} \right.    
\end{displaymath}
where $t_{\rm sweep}$ is determined by when $R_s(t) = R_{d}$. 
These equations can be used with the approximate light curve integral equation~(\ref{eq:lcintegral}) to estimate the properties of various 
disk and ejecta parameters.

\acknowledgments

This work was supported in part by the Theoretical Astrophysics Center at UC Berkeley. 
DK is supported in part by a Department of Energy Office of Nuclear
Physics Early Career Award, and by the Director, Office of Energy
Research, Office of High Energy and Nuclear Physics, Divisions of
Nuclear Physics, of the U.S. Department of Energy under Contract No.
DE-AC02-05CH11231. 
 Numerical computations utilized the Savio computational cluster resource provided by the Berkeley Research Computing program at the University of California, Berkeley (supported by the UC Berkeley Chancellor, Vice Chancellor of Research, and Office of the CIO).

\bibliographystyle{apj} 
\bibliography{msbib}

\end{document}